\documentclass[journal]{IEEEtran}

\usepackage{blindtext}
\usepackage{graphicx}
\usepackage{amsmath}
\usepackage{mathpazo}	
\usepackage{tikz}
\usepackage{color}
\usepackage{cite}
\usepackage{subfig}
\usepackage{wrapfig}
\usepackage{graphicx}

\newcommand{\fjv}[1]{{\color{red}{\textnormal{#1}}}}

\ifCLASSINFOpdf
\else
\fi
\begin{document}
%
\title{Transceiver Noise Characterization based on Perturbations}
%
%
%

\author{{
    F.J. Vaquero-Caballero,     
	D.J. Ives,
 	S.J. Savory
}


%
%


\thanks{FJVC (e-mail: fjv24@cam.ac.uk), DJI, and SJS are the Electrical Engineering Division, Department of Engineering, University of Cambridge, Cambridge CB3 0FA, UK.}	
\thanks{Manuscript received January 25, 2021; }}    

%
%

\markboth{Journal of Lightwave Technology}%
{Shell \MakeLowercase{\textit{et al.}}: Bare Demo of IEEEtran.cls for IEEE Journals}
%



\maketitle

\begin{abstract}
In this paper, we discuss a new technique for measuring transceiver noise, skew, and for the detection of uncompensated transceiver impairments. The method consists on the introduction of frequency domain notch or notches at the transmitter, which allows to estimate of the Signal-to-Noise and Distortion Ratio (SNDR) at different stages over the transmission chain. The proposed technique requires the detection of the signal spectrum using either: a modem receiver, an Optical Spectrum Analyzer (OSA), an oscilloscope, or an Electrical Spectrum Analyzer (ESA), depending on the stage where the spectrum is measured and requiring minimal processing of the received signal.

Our analysis is focused on Ciena's WaveLogic Ai commercial transceiver and a 95 Gbaud DAC. We demonstrate that symmetrically placed frequency domain notches can be used to mitiguate the influence of crosstalk on the SNDR estimates. 

Finally, we show how a single notch spectrum can be used to detect and compensate for impairments, and we perform skew estimation as an example.
\end{abstract}


%
\IEEEpeerreviewmaketitle

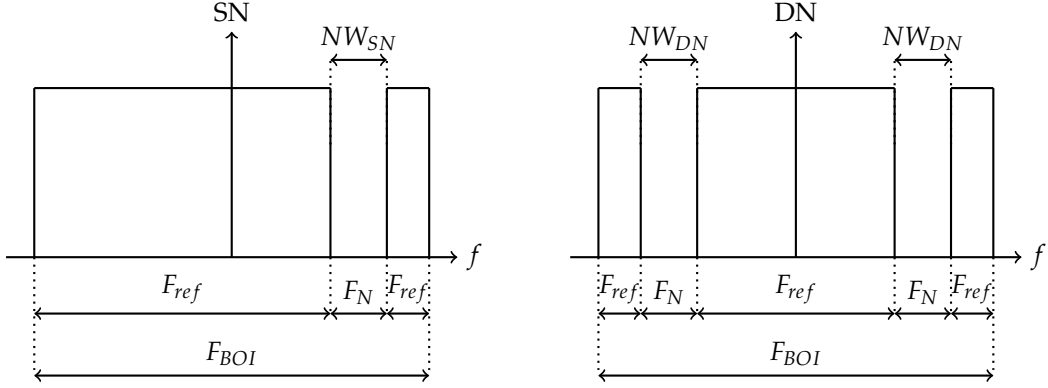
\begin{figure*}[h!]
\centering
\begin{tikzpicture}[scale=1.50]
\newcommand\Hydra[3]{
\draw [->,thick]	(-2+#1,0+#2) 	    -- (2+#1,0+#2)node[right=-0.1] {$f$};	
\draw [thick]   	(-1.75+#1,0+#2)    	-- (-1.75+#1,1.5+#2);	
\draw [->,thick]	(#1,0+#2)           -- (#1,2+#2) node[above] {#3};
\draw [thick] 		(1.75+#1,#2) 	    -- (1.75+#1,1.5+#2);		

\draw [thick] 		(-1.75+#1,1.5+#2)	-- (-1.375+#1,1.5+#2);
\draw [thick] 		(-0.875+#1,#2) 		-- (-0.875+#1,1.5+#2);
\draw [thick] 		(-1.375+#1,0+#2)   	-- (-1.375+#1,1.5+#2);
\draw [thick] 		(-0.875+#1,1.5+#2)	-- (0.875+#1,1.5+#2);
\draw [thick] 		(1.375+#1,0+#2)   	-- (1.375+#1,1.5+#2);	
\draw [thick] 		(0.875+#1,0+#2)  	-- (0.875+#1,1.5+#2);	
\draw [thick] 		(1.375+#1,1.5+#2)	--  (1.75+#1,1.5+#2);
\draw [thick,<->] 	(0.25-2+#1,-0.675+0.125+#2-0.5)--	node[above] {$F_{BOI}$} (3.75-2+#1,-0.675+0.125+#2-0.5);
\draw [thick, dotted]   (3.75-2+#1,0+#2)	--	(3.75-2+#1,0+#2-1);
\draw [thick, dotted]   (0.25-2+#1,0+#2)    --	(0.25-2+#1,0+#2-1);
\draw [thick,<->] (0.25-2+#1,#2-0.5) 		-- 	node[above] {$F_{ref}$} (-1.375+#1,#2-0.5);
\draw [thick, dotted]   (-1.375+#1,0+#2)	--	(-1.375+#1,#2-0.5);
\draw [thick, dotted]   (-0.875+#1,0+#2)	--	(-0.875+#1,#2-0.5);
\draw [thick,<->] (-1.375+#1,#2-0.5)        -- 	node[above] {$F_{N}$}(-0.875+#1,#2-0.5);

\draw [thick,<->] (-0.875+#1,#2-0.5) 		-- 	node[above] {$F_{ref}$} (2.875-2+#1,#2-0.5);
\draw [thick,<->] (3.375-2+#1,#2-0.5) 		-- 	node[above] {$F_{ref}$} (3.75-2+#1,#2-0.5);
\draw [thick,<->] (2.875-2+#1,#2-0.5)       -- 	node[above] {$F_{N}$}(3.375-2+#1,#2-0.5);
\draw [thick, dotted]   (2.875-2+#1,0+#2)	--	(2.875-2+#1,#2-0.5);
\draw [thick, dotted]   (3.375-2+#1,0+#2)   --	(3.375-2+#1,#2-0.5);
}

\newcommand\Moses[3]{
\draw [->,thick] 	(0-2+#1,0+#2) 	    -- 	(4-2+#1,0+#2)node[right=-0.1] {$f$};	
\draw [thick]   	(0.25-2+#1,0+#2)    -- 	(0.25-2+#1,1.5+#2);			    		
\draw [->,thick]	(#1,0+#2)           -- 	(#1,2+#2) node[above]{#3};    			
\draw [thick] 		(3.75-2+#1,0+#2) 	-- 	(3.75-2+#1,1.5+#2);						
\draw [thick] 		(0.25-2+#1,1.5+#2)	-- 	(2.875-2+#1,1.5+#2);				    
\draw [thick] 		(3.375-2+#1,0+#2) 	-- 	(3.375-2+#1,1.5+#2);				    
\draw [thick] 		(2.875-2+#1,0+#2)  	-- 	(2.875-2+#1,1.5+#2);				    
\draw [thick] 		(3.375-2+#1,1.5+#2)	--	(3.75-2+#1,1.5+#2); 					
\draw [thick,<->] 	(0.25-2+#1,-0.675+0.125+#2-0.5)	--	node[above] {$F_{BOI}$} (3.75-2+#1,-0.675+0.125+#2-0.5);
\draw [thick, dotted]   (3.75-2+#1,0+#2)	--	(3.75-2+#1,0+#2-1);
\draw [thick, dotted]   (0.25-2+#1,0+#2)    --	(0.25-2+#1,0+#2-1);
\draw [thick,<->] (0.25-2+#1,#2-0.5) 		-- 	node[above] {$F_{ref}$} (2.875-2+#1,#2-0.5);
\draw [thick,<->] (3.375-2+#1,#2-0.5) 		-- 	node[above] {$F_{ref}$} (3.75-2+#1,#2-0.5);
\draw [thick,<->] (2.875-2+#1,#2-0.5)       -- 	node[above] {$F_{N}$}(3.375-2+#1,#2-0.5);
\draw [thick, dotted]   (2.875-2+#1,0+#2)	--	(2.875-2+#1,#2-0.5);
\draw [thick, dotted]   (3.375-2+#1,0+#2)   --	(3.375-2+#1,#2-0.5);
}
\Hydra{5}{0}{DN}
\Moses{0}{0}{SN}

\draw [thick,<->] 	 (5.875,1.75)	-- node[above] {$NW_{DN}$} (6.375,1.75);	%
\draw [dotted,thick] (5.875,1)		-- (5.875,1.75);	
\draw [dotted,thick] (6.375,1)		-- (6.375,1.75);	

\draw [thick,<->] 	 (5.875-2.25,1.75)	-- node[above] {$NW_{DN}$} (6.375-2.25,1.75);	%
\draw [dotted,thick] (5.875-2.25,1)		-- (5.875-2.25,1.75);	
\draw [dotted,thick] (6.375-2.25,1)		-- (6.375-2.25,1.75);	

\draw [thick,<->] 	 (5.875-5,1.75)	-- node[above] {$NW_{SN}$} (6.375-5,1.75);	%
\draw [dotted,thick] (5.875-5,1)	-- (5.875-5,1.75);	
\draw [dotted,thick] (6.375-5,1)	-- (6.375-5,1.75);	
\end{tikzpicture}
\caption{Single (SN) and Dual (DN) Notch Perturbations.}
\label{fig:SingleDualNexample}
\end{figure*}

\section{Introduction}
Optical communication relies on transceivers for the generation and reception of data-carrying optical signals. Their specification and capabilities have an essential effect on the achievable capacity of optical links. Thus, the study, calibration, and bench-marking of transceivers is a relevant area of study in optical communications to provide insights into transceivers' physical characteristics and detect uncompensated impairments.

Additionally, the increasing disaggregation of optical networks into open and inter-operable equipment driving the proliferation of standards such as 400ZR, or openROADM \cite{Filer2019}, calls for ways of partitioning noise contributions by cause to support interoperability. These specifications define hardware components and their expected performance. Standard measurement methodologies are essential to ensure coherent design criteria within all vendors and manufacturers, compatibility, and compliance with standards.

A commercial coherent transmitter comprises several elements \cite{Savory2010a,Sun2020,Laperle2014,Lal2020,Kuschnerov2009,Roberts2009,Roberts2017} whose behavior can be characterized, Digital Signal Processing (DSP): FEC encoding the information stream, pulse shaping and pre-compensating for Chromatic Dispersion (CD); the Digital to Analog converters (DAC): transforming the digital electrical signal into an analog one, analog radio frequency (RF) and Electrical/Optical (E/O) modules: converting the electrical signal into an optical one at the carrier frequency, and the low internal noise optical amplifier module providing the desired output power.

The received signal is converted into the electrical domain by the analog RF and Optical/Electrical (O/E) modules to be converted again into a digital signal by the Analog to Digital Converters (ADC). Finally, a DSP processing module compensates for propagation impairments and uncompensated CD, demodulates the signal and obtains the error-free data after the Forward Error Correction (FEC) module.

Noise estimation and component characterization techniques have been studied by the research community and companies interested in bench-marking and noise budgeting \cite{Goldfarb2014,Aouini2020}. Current research is being carried out especially in phase noise measurements\cite{Ju2017,AlQadi2020}. In this introduction, we present new techniques applicable to transceiver calibration.

The eye-closure ($EC$) transceiver characterization \cite{Shiner2020} is a \fjv{Noise-to-Signal Ratio} first-order polynomial fit where the received is loaded with external white noise, typically Amplified Spontaneous Emission (ASE) source. Thus, the receiver NSR ($NSR_{RX}$) usually calculated from the pre-FEC BER can be represented as:
\begin{equation}
NSR_{RX} = \frac{NSR_{TRX} + NSR_{ASE}}{EC},
\end{equation}
where the Amplified Spontaneous Emission (ASE) Noise-to-Signal Ratio is $NSR_{ASE}$, monitored by an Optical Spectrum Analyzer (OSA). $NSR_{RX}$ is obtained from the reported pre-FEC BER of the card. Thus, the channel noise contributions divided by an EC resulting in an effective NSR that determines the pre-FEC BER (typically EC $<$ 1). \fjv{Thus, the eye-closure captures the total penalty enhancement in the receiver over the loaded ASE noise as a result of imperfections of the transceiver and its processing.}

\fjv{Moreover, the transceiver amplitude response} can be measured by loading complex white Gaussian noise as \fjv{the transmitted instruction} into the card and capturing the spectrum with a high-resolution OSA to observe the transmitter's induced frequency response, \fjv{in this paper we consider OSA resolutions of 150 MHz}. This measurement can be performed on any combination of I and Q on one or both polarizations. \fjv{Since the loaded white Gaussian noise instruction at the transmitter is flat, the exhibited frequency dependence of the transmitted spectrum at the OSA is a result of the transmitter's frequency response.}

Similarly, the receiver amplitude response can be estimated from an complex white Gaussian noise as input, \fjv{sharing the same first principles compared to the transmitter amplitude response estimation methodology}. Transmitter I-Q skew can be calibrated by loading the same signal with different skew compensations, and measuring the flatness of the resultant spectrum. \fjv{the transmission of the same signal in I and Q results in constructive and destructive interferences that creates an undulatory spectrum,} an un-skewed condition results in a maximally flat spectrum.

Despite the existing methodologies, there is still a need for efficient and straightforward methods to estimate transceiver noise contributions.

In this paper, we investigate a novel method for estimating noise contributions at different stages of an optical transceiver outfitted with digital-to-analog and analog-to-digital converters (DACs and ADCs) in the transmitter and receiver, respectively. A set of spectral perturbations are introduced into the transmitter to measure the underlying noise floor segments that are spectrally stitched with contiguous noise floor spectra to separate the frequency domain signal and noise components, and an associated Signal-to-Noise-Distortion Ratio (SNDR) is obtained.

While similar approaches have been considered for wireless communications and ADC characterization \cite{HankZumbahlen2008,Irons2000a}, to the best of our knowledge, this represents the first paper exploring applying perturbations for the characterization of optical transceivers. The proposed approach has already been used for calibration, noise discrimination and bench-marking of transceivers in a manufacturing environment.

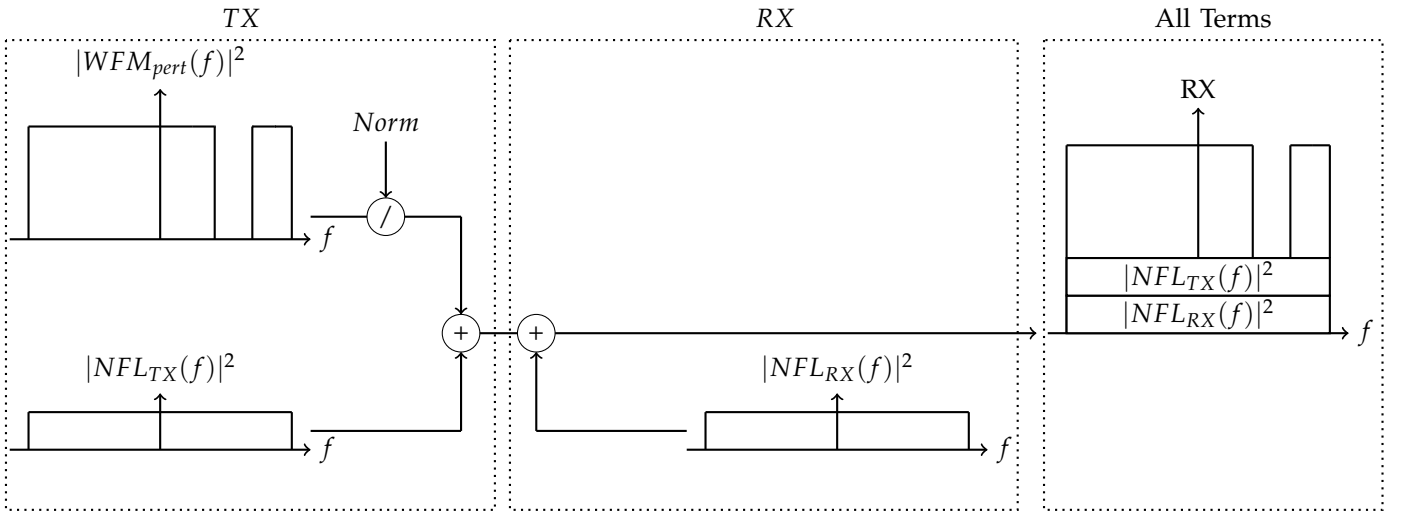
\begin{figure*}[ht]
\centering
\begin{tikzpicture}
\centering
\newcommand\NoiseF[3]{  
\draw [->,thick] (-2+#1,-0.25+#2)   -- (2.0+#1,-0.25+#2) node[right=-0.1] {$f$};
\draw [thick] (-1.75+#1,-0.25+#2)   -- (-1.75+#1,0.25+#2);					    
\draw [->,thick] (0+#1,-0.25+#2)    -- (0+#1,0.5+#2) node[above] {#3};	        
\draw [thick] (1.75+#1,-0.25+#2)    -- (1.75+#1,0.25+#2);					    
\draw [thick] (-1.75+#1,0.25+#2)    -- (1.75+#1,0.25+#2);				        
}

\newcommand\StackOfNoises[3]{  
\draw [->,thick] (-2+#1,-0.25+#2)   -- (2.0+#1,-0.25+#2) node[right=-0.1] {$f$};
\draw [thick] (-1.75+#1,-0.25+#2)   -- (-1.75+#1,0.25+#2);					    
\draw [->,thick] (0+#1,-0.25+#2)    -- (0+#1,0.5+#2) node[above] {#3};	        
\draw [thick] (1.75+#1,-0.25+#2)    -- (1.75+#1,0.25+#2);					    
\draw [thick] (-1.75+#1,0.25+#2)    -- (1.75+#1,0.25+#2);				        
}
\newcommand\WFMf[3]{
\draw [->,thick] (0-2+#1,#2)	-- (4-2+#1,#2)node[right=-0.1] {$f$};	
\draw [thick]   (0.25-2+#1,#2)	-- (0.25-2+#1,1.5+#2);			    
\draw [->,thick] (#1,#2)    	-- (#1,2+#2) node[above] {#3};
\draw [thick] (1.75+#1,#2) 	    -- (1.75+#1,1.5+#2);					
\draw [thick] (0.25-2+#1,1.5+#2)-- (0.875+#1,1.5+#2);				
\draw [thick] (1.375+#1,#2)   	-- (1.375+#1,1.5+#2);				
\draw [thick] (0.875+#1,#2)  	-- (0.875+#1,1.5+#2);				
\draw [thick] (1.375+#1,1.5+#2)	--  (1.75+#1,1.5+#2); 				
\draw [thick,<->] (0.25-2+#1,-0.675+0.125+#2)	--	node[above] {$F_{BOI}$} (3.75-2+#1,-0.675+0.125+#2);
}

\newcommand\WFMfppnp[3]{
\draw [->,thick] (0-2+#1,#2)	-- (4-2+#1,#2)node[right=-0.1] {$f$};		
\draw [thick]   (0.25-2+#1,#2)	-- (0.25-2+#1,1.5+#2);			    		
\draw [->,thick] (#1,#2)    	-- (#1,2+#2) node[above] {#3}; 				
\draw [thick] (1.75+#1,#2) 	    -- (1.75+#1,1.5+#2);						
\draw [thick] (0.25-2+#1,1.5+#2)-- (0.875+#1-0.45,1.5+#2);					
\draw [thick]	(0.875+#1-0.15,1.5+#2)		--	(0.875+#1-0.15,1.5+#2);		
\draw [thick]	(0.875+#1-0.45,1.5+#2)		--	(0.875+#1-0.15,1.5+#2);		
\draw [thick]	(1.375+#1-0.15,1.5+#2)		--	(1.375+#1-0.15,1.5+#2);		
\draw [thick]	(1.375+#1+0.15,1.5+#2)		--	(1.375+#1+0.15,1.5+#2);		
\draw [thick]	(1.375+#1-0.15,1.5+#2)		--	(1.375+#1+0.15,1.5+#2);		
\draw [thick] (1.375+#1+0.15,1.5+#2)		--  (1.75+#1,1.5+#2); 			
\draw [thick] (1.375+#1-0.15,#2)			--	(1.375+#1-0.15,1.5+#2);		
\draw [thick] (0.875+#1-0.15,#2)			--	(0.875+#1-0.15,1.5+#2);		

%
%


}

\newcommand\WFMfppnpEdited[3]{
\draw [->,thick] (0-2+#1,#2)	-- (4-2+#1,#2)node[right=-0.1] {$f$};		    
\draw [thick]   (0.25-2+#1,#2)	-- (0.25-2+#1,1.5+#2+1);			    		
\draw [->,thick] (#1,#2)    	-- (#1,2+#2+1
) node[above] {#3}; 				
\draw [thick] (1.75+#1,#2+1) 	    -- (1.75+#1,1.5+#2+1);						
\draw [thick] (0.25-2+#1,1.5+#2+1)-- (0.875+#1-0.15,1.5+#2+1);					
\draw [thick]	(1.375+#1-0.15,1.5+#2+1)	--	(1.375+#1+0.15,1.5+#2+1);	    
\draw [thick] (1.375+#1+0.15,1.5+#2+1)		--  (1.75+#1,1.5+#2+1); 		    
\draw [thick] (1.375+#1-0.15,#2+1)			--	(1.375+#1-0.15,1.5+#2+1);		
\draw [thick] (0.875+#1-0.15,#2+1)			--	(0.875+#1-0.15,1.5+#2+1);		

	%
	%
	%
\draw	[thick,fill=white] (.25-2+#1,#2)	rectangle	node{$|NFL_{RX}(f)|^2$}	(3.75-2+#1,#2+0.5);
\draw	[thick,fill=white] (.25-2+#1,#2+0.5)rectangle	node{$|NFL_{TX}(f)|^2$}	(3.75-2+#1,#2+1);

}

\WFMfppnp{0}{0.7}{$|WFM_{pert}(f)|^2$}
\NoiseF{0}{-1.85}{$|NFL_{TX}(f)|^2$}
\draw 	[thick] (2,1) 		-- (4,1);
\draw 	[thick] (2,-1.85) 	-- (4,-1.85);
\draw	[thick, dotted] (-2.05,3.35) 	rectangle ++(6.5,-6.25);	
\draw 	(0.575+0.5,3.65) node {$TX$};
\filldraw [draw,fill=white]	(4,-0.55)	circle (0.25) node {+};
\draw 	[thick,->] (4,1) 			--	(4,-0.55+0.25);

\filldraw [draw,fill=white]	(3,1) 		circle (0.25) node {/};
\draw 	[thick,<-] (3,1.25)			--	(3,2) node[above]{$Norm$};

\draw 	[thick,->] (4,-1.85)		--	(4,-0.55-0.25);
\draw 	[thick,->] (3.25+1,-0.55)	-- (3.75+2-0.75,-0.55);
\filldraw [draw,fill=white]	(3.75+2-0.75,-0.55)	circle (0.25) node {+};	
	
\draw 	[thick,->] 	(3.75+2-0.75,-1.85)	--	(3.75+2-0.75,-0.55-0.25);
\draw 	[thick] 	(3.75+2-0.75,-1.85)	--	(3.75+4-0.75,-1.85);
\draw 	(9.375-1.2,3.65) node {$RX$};

\NoiseF{3.75+6-0.75}{-1.85}{$|NFL_{RX}(f)|^2$}
\draw	[thick, dotted] (11.75,3.35) 	rectangle ++(4.5,-6.25);
\draw 	(14,3.65) node {All Terms};

\draw 	[thick,->] (4+2-0.75,-0.55) 		--	(3.75+8-0.1,-0.55);
\draw	[thick, dotted] (5.4-0.75,3.35) 	rectangle ++(6.75,-6.25);	
\WFMfppnpEdited{3.75+9+2-0.95}{-0.55}{RX}

\end{tikzpicture}
\caption{Transmission model considering transmitter (TX) with the normalization loop \fjv{(Norm)}, the receiver (RX) noises, and the totality of noises added.}
\label{fig:FullTransModel}
\end{figure*}
\section{Definition of a Perturbation}
Spectral perturbations can be loaded into the transmitter filter responsible of pre-compensating chromatic dispersion, or directly performed to the waveform (WFM) instruction. Such perturbations can be modeled as a real frequency domain filter over the transmitted field, $H_{pert}(f)$, which can enhance some frequency components, deplete others, or keep them unchanged:
\begin{equation}
WFM_{pert}(f)=WFM_{org}(f) H_{pert}(f),
\label{eq:PertDef}
\end{equation}
\begin{equation*}
H_{pert}(f) = \begin{cases}
\sqrt{G(A)}, &  f \in F_A\\
\sqrt{G(B)}, &  f \in F_B\\
1, 		   &  f \in F_U \\
\end{cases},
\end{equation*}

where: $WFM_{pert}(f)$, $WFM_{org}(f)$, and $H_{pert}(f)$ are denoted in frequency domain. $F_A$, $F_B$ are arbitrary frequency regions of the spectrum whose associated gains are $G(A)$ and $G(B)$; $F_{U}$ corresponds to the unperturbed region of the spectrum whose gain is 1, in this case we will refer to that region as $F_{ref}$.

In this application, \fjv{the gain of the perturbation is set to 0, $G=0$ or $G=-\infty$ [dB], fully attenuating} the DAC instruction's power spectral density. Such perturbations are denoted as Notches (N), and $F_N$ is their frequency range.

Two types of perturbations are considered in this paper, as illustrated in Figure \ref{fig:SingleDualNexample}: a single (SN) and a dual notch (DN) perturbation. For this specific application, dual notches filters are symmetrical w.r.t. the carrier. $NW$ denotes the width of each notch. The bandwidth of interest (BOI) specifies the frequency range of our transmitted signal. 

\section{Noise Floor Measurement}
\begin{figure*}[ht]
\centering
\begin{tikzpicture}
\newcommand\WFMf[6]{
\draw [->,thick](0-2+#1,#2)--(4-2+#1,#2)node[right=-0.1]{$f$};
\draw [->,thick](#1,#2)--(#1,2+#2+#4+0.85)node[above]{#3};

\def\sqstr{-1.75}
\def\argiii{#5}
\ifx\argiii\sqstr
\else
  	\draw [thick]	(0.25-2+#1,#2+#4)	-- (0.25-2+#1,1.5+#2+#4);
	\draw [thick] 	(#5+#1,#2+#4)  		-- (#5+#1,1.5+#2+#4);
\fi

\def\sqstr{1.75}
\def\argiii{#6}
\ifx\argiii\sqstr
\else
	\draw [thick] (1.75+#1,#2+#4)		-- (1.75+#1,1.5+#2+#4);
	\draw [thick] (#6+#1,#2+#4) 		-- (#6+#1,1.5+#2+#4);
\fi

\draw [thick] (0.25-2+#1,1.5+#2+#4)	-- (#5+#1,1.5+#2+#4);
\draw [thick] (#6+#1,1.5+#2+#4)		-- (1.75+#1,1.5+#2+#4);

\draw [thick,dashed](0.25-2+#1,#2+#4)--(0.25-2+#1,1.5+#2+#4+0.75);
\draw [thick,dashed](0.25-2+#1+1.167,#2+#4)--(0.25-2+#1+1.167,1.5+#2+#4+0.75);
\draw [thick,dashed](1.75+#1-1.167,#2+#4)--(1.75+#1-1.167,1.5+#2+#4+0.75);
\draw [thick,dashed](1.75+#1,#2+#4)--(1.75+#1,1.5+#2+#4+0.75);

\draw [thick,<->](0.25-2+#1,1.5+#2+#4+0.375)--node[above]{$F_1$}(0.25-2+#1+1.167,1.5+#2+#4+0.375);

\draw [thick,<->](0.25-2+#1+1.167,1.5+#2+#4+0.375)--node[above,xshift=7pt]{$F_2$}(1.75+#1-1.167,1.5+#2+#4+0.375);

\draw [thick,<->](1.75+#1-1.167,1.5+#2+#4+0.375)--node[above]{$F_3$}(1.75+#1,1.5+#2+#4+0.375);

}	
\WFMf{-5}{0}{$|RX_1(f)|^2$}{0.5}{-1.75}{-1.75+1.167};
\draw[draw=black,fill=white] (-1.75-5,0.5) rectangle node{$|NFL(f)|^2$}(1.75-5,0);

\WFMf{0}{0}{$|RX_2(f)|^2$}{0.5}{-1.75+1.167}{1.75-1.167};
\draw[draw=black,fill=white] (-1.75,0.5) rectangle node{$|NFL(f)|^2$}(1.75,0);

\WFMf{5}{0}{$|RX_3(f)|^2$}{0.5}{1.75-1.167}{1.75};
\draw[draw=black,fill=white] (-1.75+5,0.5) rectangle node{$|NFL(f)|^2$}(1.75+5,0);

\end{tikzpicture}
\caption{Illustration of spectral stitching measurements to obtain the NFL of a series of measurements.}
\label{fig:NFLmeasurement}
\end{figure*}
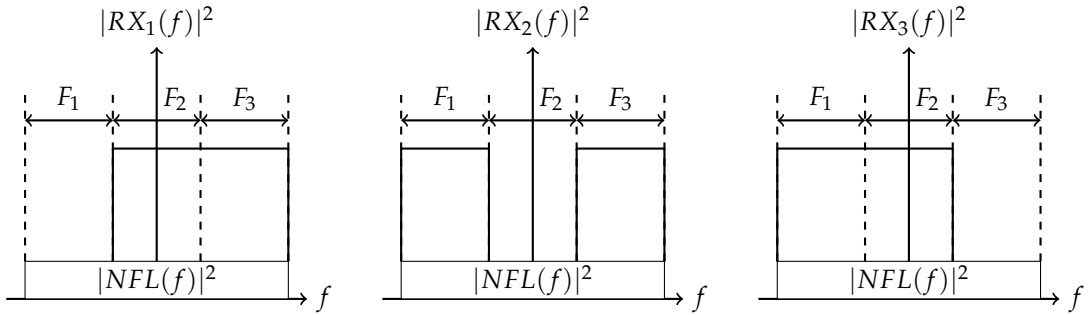

\begin{figure*}[ht]%
    	\centering
    	\subfloat[]{{\includegraphics[width=9.3cm]{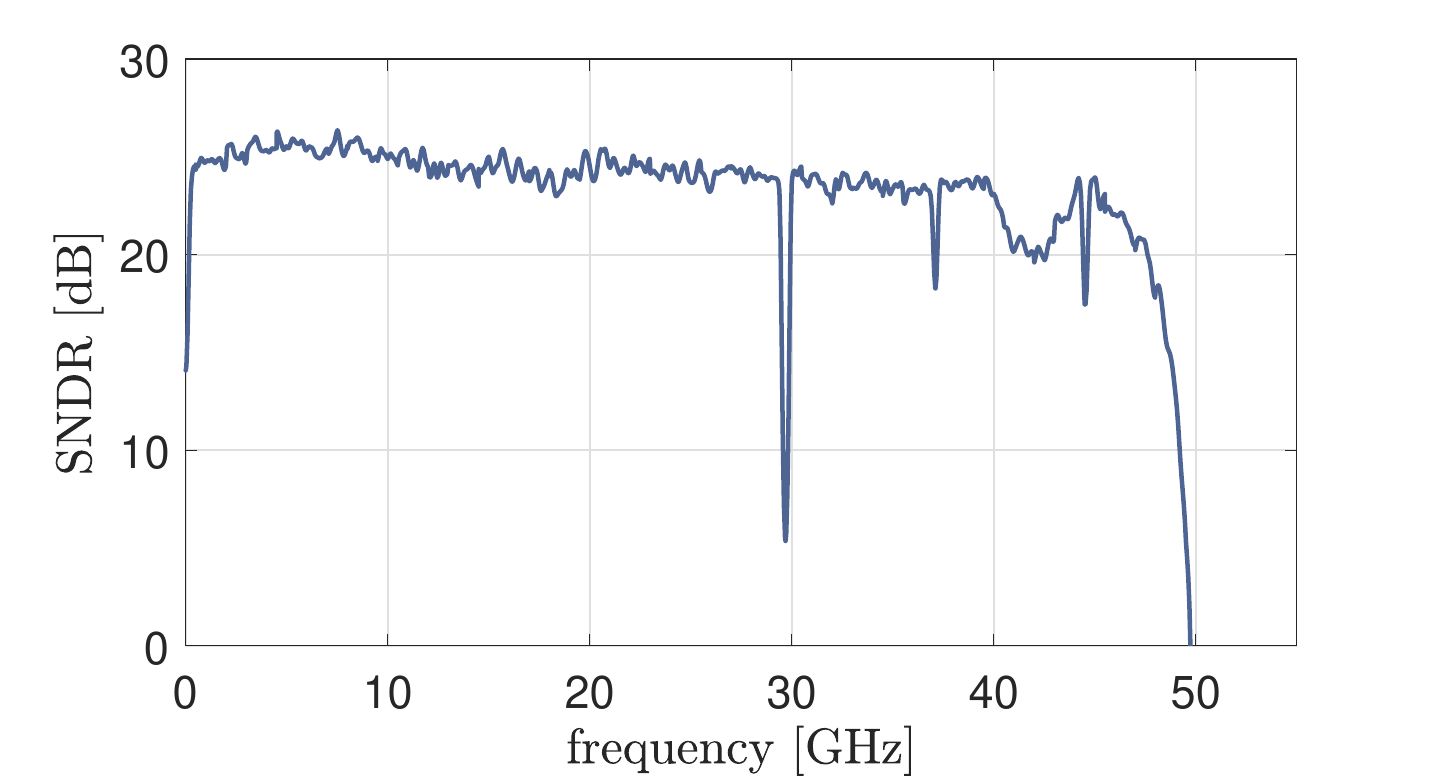}}}%
	    \subfloat[]{{\includegraphics[width=9.3cm]{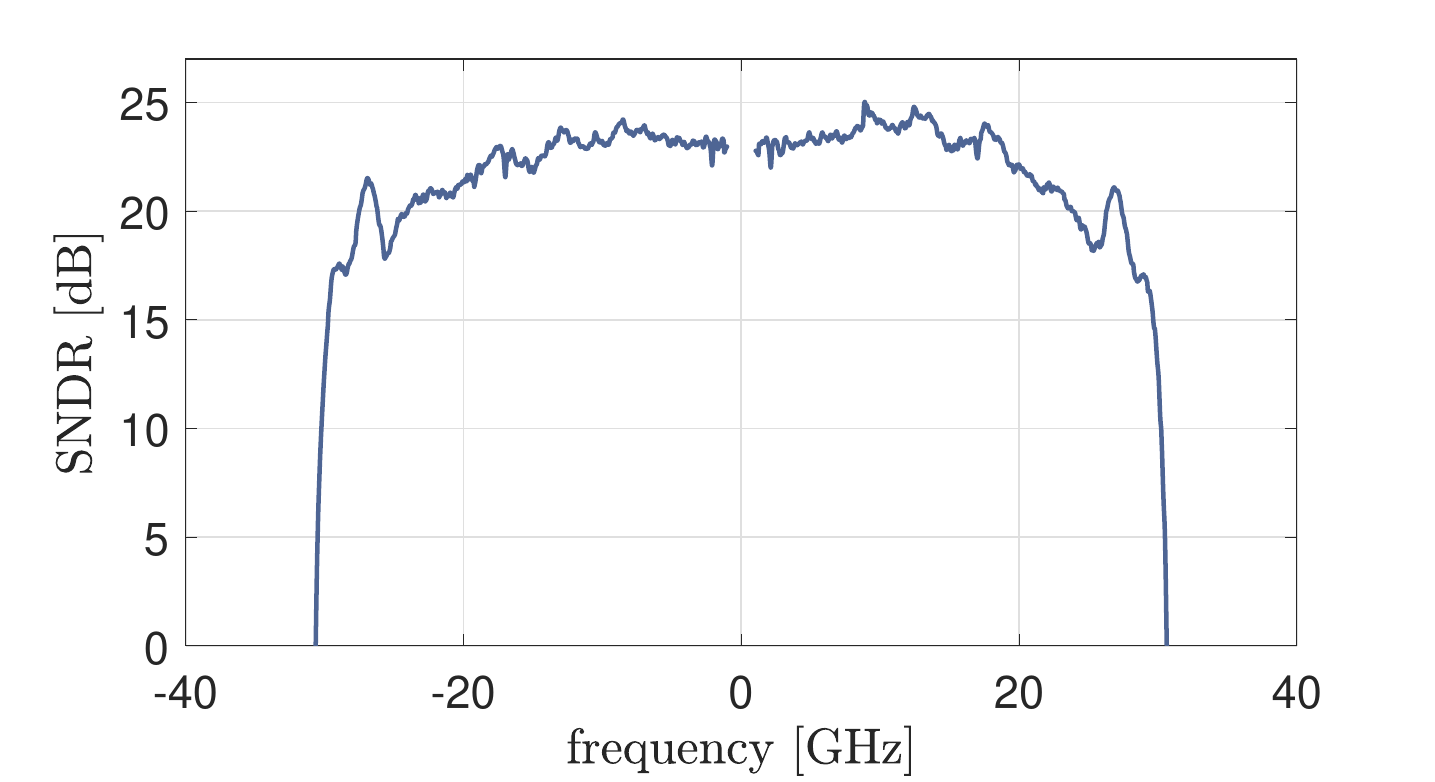}}}
    	\\
	\subfloat[]{{\includegraphics[width=9.3cm]{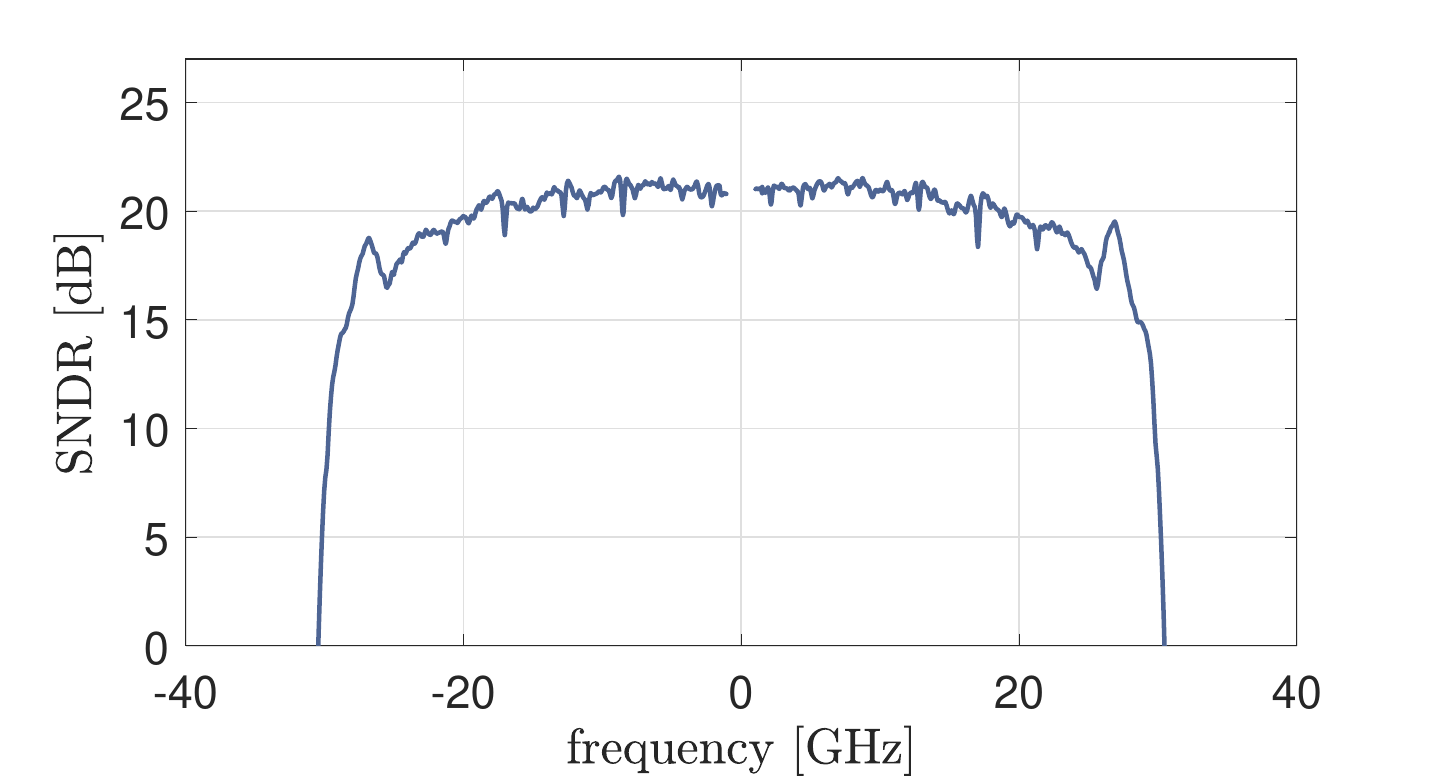}}}
	\subfloat[] {{\includegraphics[width=9.3cm]{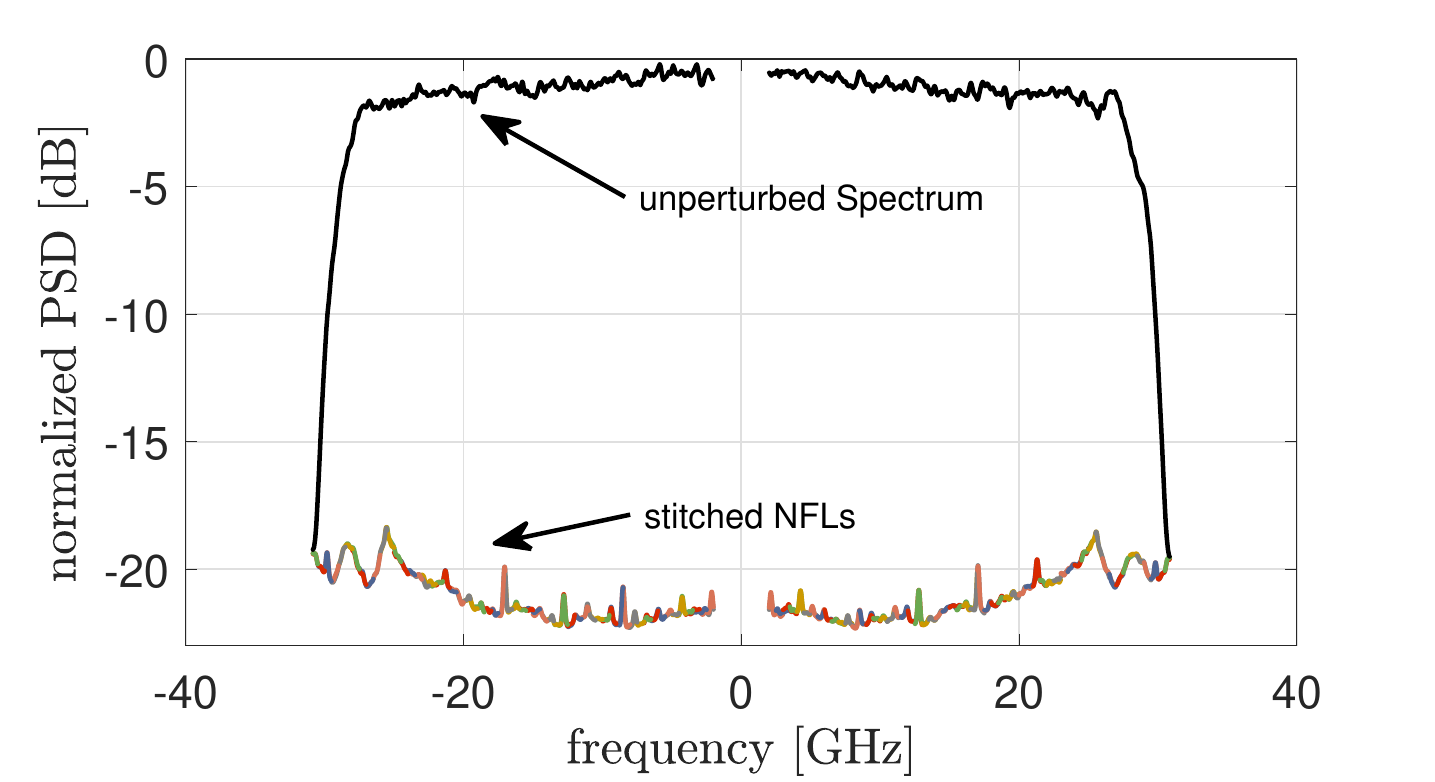}}}%
	\\
    \centering
	\caption{Different SNDR measurements, (a) E2E for a 95GBaud application, (b) card2OSA for WLAi, and (c) Card2card for WLAi. (d) corresponds to the card2card spectral stitching performed for WLAi to obtain SNDR.}
    \label{fig:NFLest}%
\end{figure*}

Figure \ref{fig:FullTransModel} presents a noise aggregation model where the perturbed waveform $WFM_{pert}$ \fjv{of the transmitter side (TX)} is followed by a normalization loop to maintain the power of $WFM_{pert}$ constant. The normalization factor is defined as: 
\begin{equation}
Norm = \frac{\int_{<F_{BOI}>} |WFM_{pert}(f)|^2 \; df}{\int_{<F_{BOI}>} |WFM_{org}(f)|^2 \; df}
\label{eq:NormTerm}
\end{equation}

A transmitter noise floor term, $NFL_{TX}$, is included to account for the noises resulted from generation from DAC, DSP, analog RF, and  E/O. The measurement at the receiver also introduces a noise floor, $NFL_{RX}$. Depending on the type of receiver considered, that contribution can be negligible (e.g.: for an OSA) or significant (e.g.: for a coherent receiver and ADC). \fjv{Since the applications of interest of this paper do not consider propagation,} any other impairments such as fiber attenuation are assumed to be negligible for this application. 

Assuming that the introduced perturbations occupy a small portion of the transmit spectrum, and the waveform peak-to-rms ratio remains unaffected, the NFL contributions can be considered constant (a peak-to-rms increment of 8\% was observed for the experiments of this section). The received signal power spectral density is:

\begin{equation}
\begin{split}
|RX_i(f)|^2 = |NFL(f)|^2, \quad f \in F_N
\\
|RX_i(f)|^2 = |NFL(f)|^2 + \frac{|WFM(f)|^2}{Norm_i}, \quad f \in (F_{BOI}-F_N)
\end{split}
\end{equation}

Figure \ref{fig:NFLmeasurement} illustrates an example where 3 spectral regions are defined $F_1$, $F_2$, and $F_3$, and 3 SN perturbations measurements are performed, \fjv{the width of the perturbation the figure is exaggerated for illustration purposes}. The frequency of the SN is changed across the BOI of the signal to expose the underlying NFL. The considered single notch or dual notch perturbations occupy a small fraction of the BOI so as to not unduly contaminate a noise floor measurement or significantly alter the operation of the transmitter. \fjv{In this paper, the $NW_{SN}$ and $NW_{DN}$ considered are}.

In the particular example of Figure \ref{fig:NFLmeasurement}, $WFM$ and $NLF$ can be calculated for $F_1$ as:
\begin{equation}
\begin{split}
|NFL(f)|^2 =  |RX_1(f)|^2, \quad f \in F_1
\\
|WFM(f)|^2 =  (|RX_2(f)|^2-|RX_1(f)|^2)Norm_2 ,\quad f \in F_1,
\end{split}
\end{equation}
\fjv{where the WFM contribution is calculated as the subtraction between the second perturbed spectra ($RX_2$) and the first perturbed spectra ($RX_1$)}

Moreover, if the Notch is small, ($Norm\approx 1$), \fjv{or there is not normalization loop on the transmitter}:
\begin{equation}
\begin{split}
|NFL(f)|^2 =  |RX_1(f)|^2, \quad f \in F_1
\\
|WFM(f)|^2 \approx  |RX_2(f)|^2-|RX_1(f)|^2 ,\quad f \in F_1,
\end{split}
\end{equation}
such approximation is not used in this paper \fjv{since the considered transmitter includes a normalization loop}.

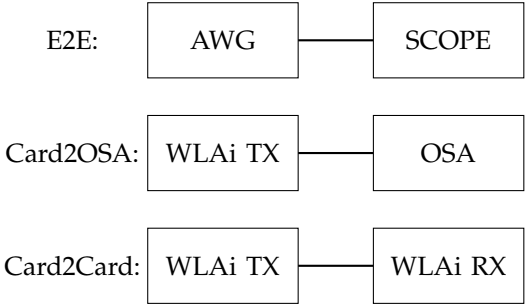
\begin{figure}[h!]
\centering
\begin{tikzpicture}
\newcommand\AMP[4]{
\draw [thick] (#1,-#4+#2)	--	(#1,#4+#2);
\draw [thick] (#1, #4+#2)	--	(#1+#3,#2);
\draw [thick] (#1,-#4+#2)	--	(#1+#3,#2);
}
\newcommand\fiber[4]{
\draw (#1+#4,#2+#3) circle (#3);
\draw (#1,#2+#3) 	circle (#3);
\draw (#1-#4,#2+#3) circle (#3);
}
\draw   [draw=black] (-1+3.5,0.5) rectangle node{AWG} (1+3.5,-0.5);
\draw   [thick,-] (4.5,0)	--	(5.5,0);
\draw   [draw=black,fill=white] (5.5,0.5) rectangle node{SCOPE} (7.5,-0.5);
\draw 	(-2+3.5,-0) node {E2E:};
\draw[draw=black] (-1+3.5,0.5-1.5) rectangle node{WLAi TX} (1+3.5,-0.5-1.5);
\draw [thick,-] (4.5,0-1.5)	--	(5.5,0-1.5);
\draw[draw=black,fill=white] (5.5,0.5-1.5) rectangle node{OSA} (7.5,-0.5-1.5);
\draw 	(-2+3.5,-1.5) node {Card2OSA:};
\draw[draw=black] (-1+3.5,0.5-3) rectangle node{WLAi TX} (1+3.5,-0.5-3);
\draw [thick,-] (4.5,0-3)	--	(5.5,0-3);
\draw[draw=black,fill=white] (5.5,0.5-3) rectangle node{WLAi RX} (7.5,-0.5-3);
\draw 	(-2+3.5,-3) node {Card2Card:};

\end{tikzpicture}
\caption{Diagram of the experimental setups.}
\label{fig:ExperimentalSetups}
\end{figure}
A measurement of this kind is compatible with many channel interfaces: for instance, for a spectral measurement with an OSA at the output of the transmitter, the NFL will only consist of $NFL_{TX}$; if it is performed at the receiver but with negligible other channel noises ($NLN+ASE\simeq 0$), the measured NFL from the perturbation technique will correspond to the addition of contributions of the transmitter and receiver up to the measurement interface.

Figure \ref{fig:NFLest} shows three experimental cases: a, b, and c, where noises are isolated according to the measurement interface. The card is connected back-to-back (B2B) with negligible ASE or propagation nonlinear noise in the present measurements, and the measurements are performed with dual notch perturbations. Thus, the following measurements are considered in this paper:

\begin{itemize}
\item Electrical-to-Electrical (E2E): Figure \ref{fig:NFLest}(a) an oscilloscope measures the perturbed spectrum to the output of a DAC. At this stage, the baseband signal can be fully represented by its positive frequencies. The generated signal is mainly affected by quantization noise, captured in $NFL_{TX}$. Additionally, it can also be measured by an Electrical Spectrum Analyzer (ESA). Depending on the type of equipment used, the noise of the measurement equipment can be considered negligible. \fjv{The SNDR at 30GHz is due to a clock artifact.}
\item Card-to-OSA  (Card2OSA): Figure \ref{fig:NFLest}(b) a high-resolution optical spectrum analyzer (OSA) is connected at the output of the transmitter. The measured spectrum includes both electrical and optical noises from the transmitter. Individual polarization contributions can also be measured by transmitting a single polarization.
\item Card-to-card (Card2card): Figure \ref{fig:NFLest}(c), the spectrum is captured by the card. Containts transmitter noise, $|NFL_{TX}(f)|^2$,  as well as optical and electrical noises up to the ADC output. Receiver DSP noises are not present since the spectrum is captured immediately after the ADC without any DSP processing.
\end{itemize}
\begin{figure*}[h!]%
    	\centering
    	\subfloat[]{{\includegraphics[width=9.3cm]{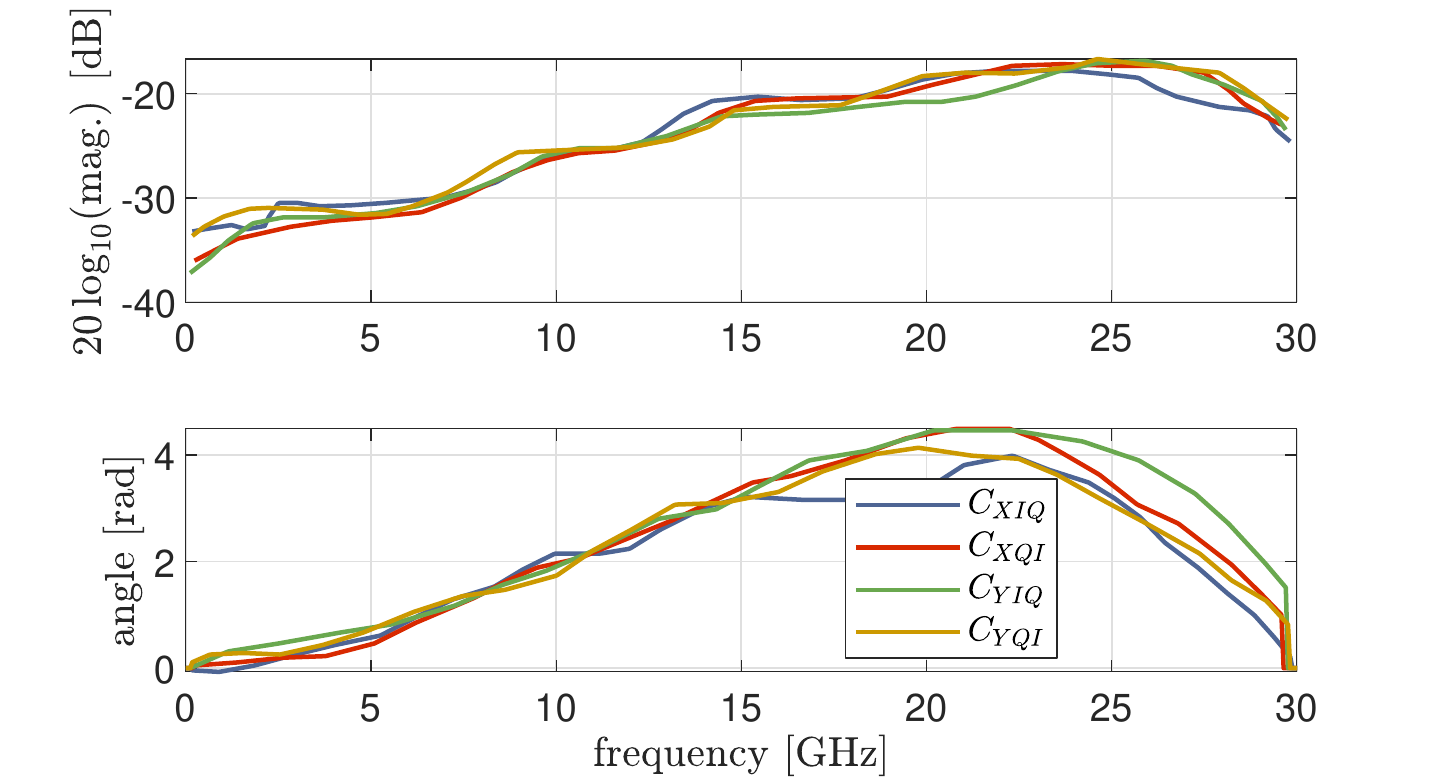}}}		%
	\subfloat[]{{\includegraphics[width=9.3cm]{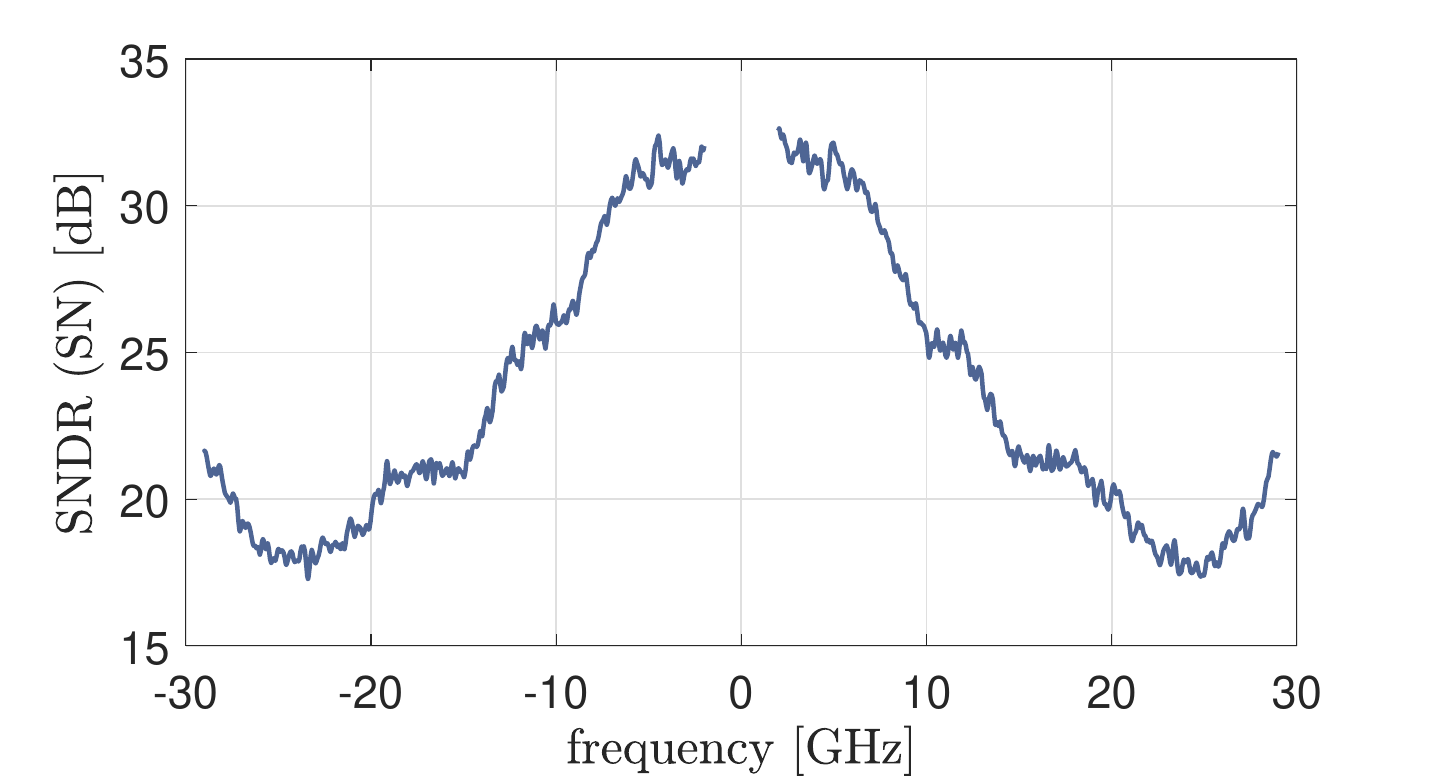}}}			%
    \centering
	\caption{Crosstalk and SNDR estimates: (a) Synthesized crosstalk, and (b) resultant SN SNDR profile from Crosstalk only.}
	\centering
	\label{fig:Xtalkest}
\end{figure*}
Figure \ref{fig:NFLest}(d), illustrates the individual NFL measurements (in color) stitched to estimate the NFL of Figure \ref{fig:NFLest}(c). Figures \ref{fig:NFLest}(b,c,d) are the same WaveLogic Ai (WLAi) transciever with a dual notch where $NW=2$ GHz. Figure \ref{fig:NFLest}(a) corresponds to a 95 GBaud with 0.05 roll-off QPSK signal generated by a Keysight M8194A AWG and detected with a UXR1104A Real-Time Oscilloscope with a Dual Notch of $NW=5$ GHz. Figure \ref{fig:ExperimentalSetups} illustrates the experimental setups considered in this paper. \fjv{The notch width considered ...}


\section{Impairments in Single Notch vs Dual Notch SNDR}
The experimental measurements of the previous sections were performed with Dual Notches. Dual Notch measurements are tolerant to unresolved IQ Crosstalk, IQ imbalance, or uncompensated transceiver phase response. Single notch and dual notch SNDRs differ if the transceiver response contains a remnant uncompensated contribution. The creation of a single notch is achieved by destructive and constructive additions of the I and Q components:
\begin{equation}
\begin{split}
X_I(f)=-jX_Q(f), \, f \in F_N
\\
X_I(f)=j X_Q(f), \, f \in F_{-_N}
\\
F_{-N} = \frac{-NW}{2}<f+NC \leq \frac{NW}{2} 
\end{split}
\label{eq:SingleNotchCondition}
\end{equation}
where $X(f)=X_I(f)+jX_Q(f)$, $I$ and $Q$ denotes the real and imaginary components of $X$. In time domain: $x_I(t)=\mathbb{R}\{x(t)\}$, $x_Q(t)=\mathbb{I}\{x(t)\}$, and $F_{-N}$ \fjv{corresponds to the opposite region of the notch region   }, $Y$ polarization accordingly. When impairments are introduced, the conditions of Equation \ref{eq:SingleNotchCondition} do not apply anymore and this results in partial constructive and destructive additions of the I and Q components. If we consider Crosstalk between polarizations negligible, the IQ Crosstalk on the X polarization is:
\begin{equation}
\hat{\textbf{X}} (f)=	
 \begin{bmatrix} \hat{X}_I(f) \\ \hat{X}_Q(f) \end{bmatrix}
=	
\begin{bmatrix}
	C_{XII}(f)&	C_{XQI}(f)\\
	C_{XIQ}	(f)&	C_{XQQ}(f)\\
   \end{bmatrix}
 \begin{bmatrix} X_I(f) \\ X_Q(f) \end{bmatrix}
\label{fig:XtalkEqu}
\end{equation}
where $C_{XII}(f)$, $C_{XQQ}(f)$ corresponds to the gain or attenuation experienced by I and Q components and $C_{XQI}(f)$, $C_{XIQ}(f)$ are the Crosstalk terms of Q to I and I to Q components, respectively. $C_{XII}(f),C_{XQQ}(f) \approx 1$ (assuming transceiver compensation is applied), $C_{XQI}(f),C_{XIQ}(f)\in \mathbb{C}$, similarly for Y polarization. Also:  $C_{Xab}(f)=C^*_{Xab}(-f); \, a,b \in \{I,Q\}$. Moreover, combining Equations \ref{eq:SingleNotchCondition} and \ref{fig:XtalkEqu}:
\begin{equation}
\begin{split}
\hat{X} (f) = j  X_I (f)\Big( C_{XQI}(f) + C_{XIQ}(f) \Big), \, f \in F_N
\end{split}
\end{equation}

Hence, differences between single notch and dual notch are associated with Crosstalk, but existing symmetries between crosstalk terms can affect the estimated SNDR. For dual notches: $X_I(f)=X_Q(f)=0, \, f \in f_N$  where $F_N$ encompasses both positive and negative frequencies belonging to the dual notch. Thus, the conditions of Equation \ref{eq:SingleNotchCondition} are respected regardless of the amount of crosstalk.

Figure \ref{fig:Xtalkest} (a) plots a synthesized crosstalk profile and Figure \ref{fig:Xtalkest} (b) illustrates the estimated SNDR based on Single Notch perturbations. The equivalent dual notch SNDR for the Crosstalk of Figure \ref{fig:Xtalkest} (a) is infinite. When second-order Crosstalk terms will result a measurable SNDR for the Dual Notch scenario. \fjv{Second-order crosstalk terms refers to noise contributions that accumulates in the $F_N$ and $F_{-N}$ regions which will also be affected by the constructive and destructive additions}.

Moreover, Dual Notch is also resilient to other linear impairments such as IQ-imbalance and skew, where constructive and destructive interferences also apply. The discrepancies between single notch and dual notch approaches are attributed to departures from Equation \ref{fig:Xtalkest} and uncompensated impairments, it may indicate that performance improvements are possible. 

\section{I-Q Skew Estimation}

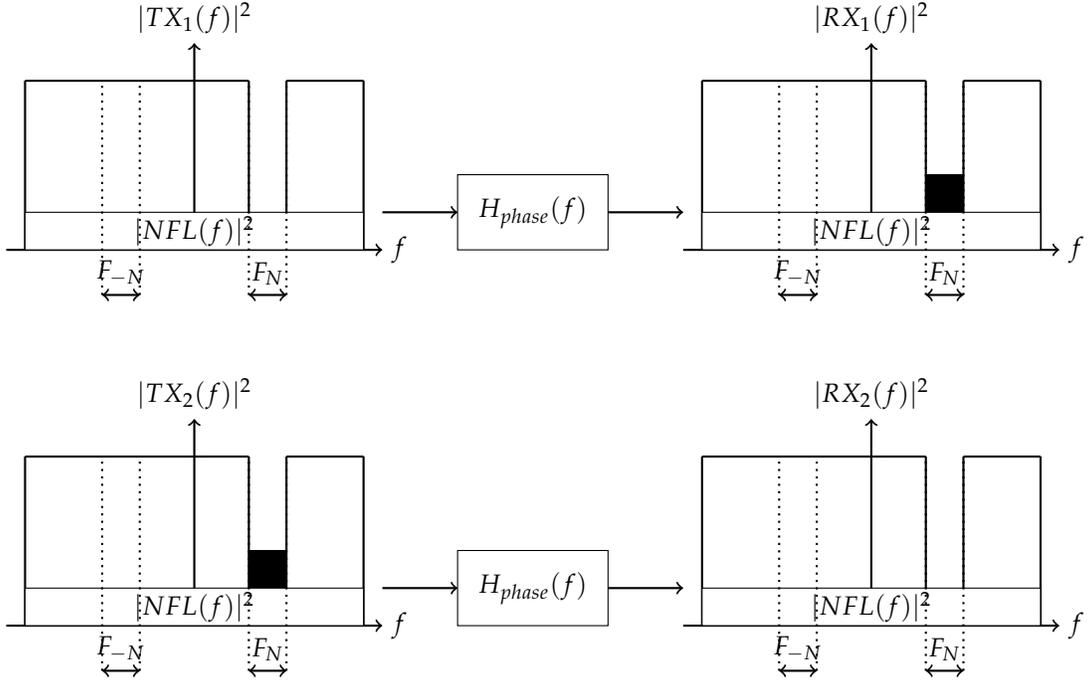
\begin{figure*}
\centering
\begin{tikzpicture}
\newcommand\WFMfppnp[3]{
%
\pgfmathsetmacro{\PPLpointR}{0.15}
\pgfmathsetmacro{\PPLpointL}{0.75}
\pgfmathsetmacro{\PPRpointL}{0.15}
\pgfmathsetmacro{\PPRpointR}{0.45}  
\pgfmathsetmacro{\WFMwidth}{2.25}   %
\pgfmathsetmacro{\AG}{0.5+0.25}   	%
\pgfmathsetmacro{\BG}{0.25+0.25}		

\draw [->,thick](-\WFMwidth-0.25+#1,#2)	-- (\WFMwidth+0.25+#1,#2)node[right=-0.1] {$f$};	
\draw [thick]   (-\WFMwidth+#1,#2)		-- (-\WFMwidth+#1,1.75+\BG+#2);			    	    
\draw [->,thick](#1,#2)    			    -- (#1,2.75+#2) node[above] {#3}; 			        
\draw [thick](\WFMwidth+#1,#2)-- (\WFMwidth+#1,1.75+\BG+#2);
\draw [thick] 	(-\WFMwidth+#1,1.75+\BG+#2)  -- (0.875+#1-\PPLpointL,1.75+#2+\BG);      
\draw [thick]	(0.875+#1-\PPLpointL,1.75+\BG+#2)   --	(0.875+#1-\PPLpointL,1.5+#2+\AG); 
\draw [thick]	(0.875+#1-\PPLpointR,1.5+#2+\AG)    --	(0.875+#1-\PPLpointR,#2);	      
\draw [thick]	(0.875+#1-\PPLpointL,1.5+#2+\AG)    --	(0.875+#1-\PPLpointR,1.5+#2+\AG); 
\draw [thick]	(1.375+#1-\PPRpointL,1.5+#2+\AG)	--	(1.375+#1-\PPRpointL,#2);         
\draw [thick]	(1.375+#1+\PPRpointR,1.5+#2+\AG)    --	(1.375+#1+\PPRpointR,1.75+\BG+#2);
\draw [thick]	(1.375+#1-\PPRpointL,1.5+#2+\AG)    --	(1.375+#1+\PPRpointR,1.5+#2+\AG); 
\draw [thick] 	(1.375+#1+\PPRpointR,1.75+\BG+#2)   --  (\WFMwidth+#1,1.75+\BG+#2);       
\draw [thick,<->] 		(0.875+#1-\PPLpointR,0.15+#2-0.75)  -- 	node[above] {$F_{N}$}	(1.375+#1-\PPRpointL,0.15+#2-0.75);
\draw [thick,<->] 		(-0.875+#1+\PPLpointR,0.15+#2-0.75)  -- 	node[above] {$F_{-N}$}	(-1.375+#1+\PPRpointL,0.15+#2-0.75);

\draw[draw=black,fill=white] (-2.25+#1,0.5+#2) rectangle node{$|NFL(f)|^2$}(2.25+#1,0+#2);
\draw [thick, dotted]	(0.875+#1-\PPLpointR,0.15+#2-0.75)--	(0.875+#1-\PPLpointR,1.75+\BG+#2);
\draw [thick, dotted]	(-0.875+#1+\PPLpointR,0.15+#2-0.75)--	(-0.875+#1+\PPLpointR,1.75+\BG+#2);
\draw [thick, dotted]	(-1.375+#1+\PPRpointL,0.15+#2-0.75)--	(-1.375+#1+\PPRpointL,1.75+\BG+#2);
\draw [thick, dotted]	(1.375+#1-\PPRpointL,0.15+#2-0.75)--	(1.375+#1-\PPRpointL,1.75+\BG+#2);
}

\newcommand\WFMfppnpB[3]{
%
\pgfmathsetmacro{\PPLpointR}{0.15}
\pgfmathsetmacro{\PPLpointL}{0.75}
\pgfmathsetmacro{\PPRpointL}{0.15}
\pgfmathsetmacro{\PPRpointR}{0.45}  
\pgfmathsetmacro{\WFMwidth}{2.25}   %
\pgfmathsetmacro{\AG}{0.5+0.25}   	%
\pgfmathsetmacro{\BG}{0.25+0.25}		

\draw [->,thick](-\WFMwidth-0.25+#1,#2)	-- (\WFMwidth+0.25+#1,#2)node[right=-0.1] {$f$};	
\draw [thick]   (-\WFMwidth+#1,#2)		-- (-\WFMwidth+#1,1.75+\BG+#2);			    	    
\draw [->,thick](#1,#2)    			    -- (#1,2.75+#2) node[above] {#3}; 			        
\draw [thick](\WFMwidth+#1,#2)-- (\WFMwidth+#1,1.75+\BG+#2);
\draw [thick] 	(-\WFMwidth+#1,1.75+\BG+#2)  -- (0.875+#1-\PPLpointL,1.75+#2+\BG);      
\draw [thick]	(0.875+#1-\PPLpointL,1.75+\BG+#2)   --	(0.875+#1-\PPLpointL,1.5+#2+\AG); 
\draw [thick]	(0.875+#1-\PPLpointR,1.5+#2+\AG)    --	(0.875+#1-\PPLpointR,#2);	      
\draw [thick]	(0.875+#1-\PPLpointL,1.5+#2+\AG)    --	(0.875+#1-\PPLpointR,1.5+#2+\AG); 
\draw [thick]	(1.375+#1-\PPRpointL,1.5+#2+\AG)	--	(1.375+#1-\PPRpointL,#2);         
\draw [thick]	(1.375+#1+\PPRpointR,1.5+#2+\AG)    --	(1.375+#1+\PPRpointR,1.75+\BG+#2);
\draw [thick]	(1.375+#1-\PPRpointL,1.5+#2+\AG)    --	(1.375+#1+\PPRpointR,1.5+#2+\AG); 
\draw [thick] 	(1.375+#1+\PPRpointR,1.75+\BG+#2)   --  (\WFMwidth+#1,1.75+\BG+#2);       
\draw [thick,<->] 		(0.875+#1-\PPLpointR,0.15+#2-0.75)  -- 	node[above] {$F_{N}$}	(1.375+#1-\PPRpointL,0.15+#2-0.75);

\draw[draw=black,fill=black] (0.875+#1-\PPLpointR,1+#2) rectangle node{}(1.375+#1-\PPRpointL,0+#2);

\draw [thick,<->] 		(-0.875+#1+\PPLpointR,0.15+#2-0.75)  -- 	node[above] {$F_{-N}$}	(-1.375+#1+\PPRpointL,0.15+#2-0.75);

\draw[draw=black,fill=white] (-2.25+#1,0.5+#2) rectangle node{$|NFL(f)|^2$}(2.25+#1,0+#2);
\draw [thick, dotted]	(0.875+#1-\PPLpointR,0.15+#2-0.75)--	(0.875+#1-\PPLpointR,1.75+\BG+#2);
\draw [thick, dotted]	(-0.875+#1+\PPLpointR,0.15+#2-0.75)--	(-0.875+#1+\PPLpointR,1.75+\BG+#2);
\draw [thick, dotted]	(-1.375+#1+\PPRpointL,0.15+#2-0.75)--	(-1.375+#1+\PPRpointL,1.75+\BG+#2);
\draw [thick, dotted]	(1.375+#1-\PPRpointL,0.15+#2-0.75)--	(1.375+#1-\PPRpointL,1.75+\BG+#2);
}

\WFMfppnp{1}{0}{$|TX_1(f)|^2$};
\draw [->,thick] (3.5,0.5)--(4.5,0.5);
\draw[draw=black,fill=white] (4.5,1) rectangle node{$H_{phase}(f)$}(6.5,0);
\draw [->,thick] (6.5,0.5)--	(7.5,0.5);
\WFMfppnpB{10}{0}{$|RX_1(f)|^2$};

\WFMfppnpB{1}{0-5}{$|TX_2(f)|^2$};
\draw [->,thick] (3.5,0.5-5)--(4.5,0.5-5);
\draw[draw=black,fill=white] (4.5,1-5) rectangle node{$H_{phase}(f)$}(6.5,0-5);
\draw [->,thick] (6.5,0.5-5)--	(7.5,0.5-5);
\WFMfppnp{10}{-5}{$|RX_2(f)|^2$};
\end{tikzpicture}
\caption{Illustration of the principle for skew estimation.}
\label{fig:PhaseEst}
\end{figure*}

\begin{figure}
\centering
\includegraphics[width=9.5cm]{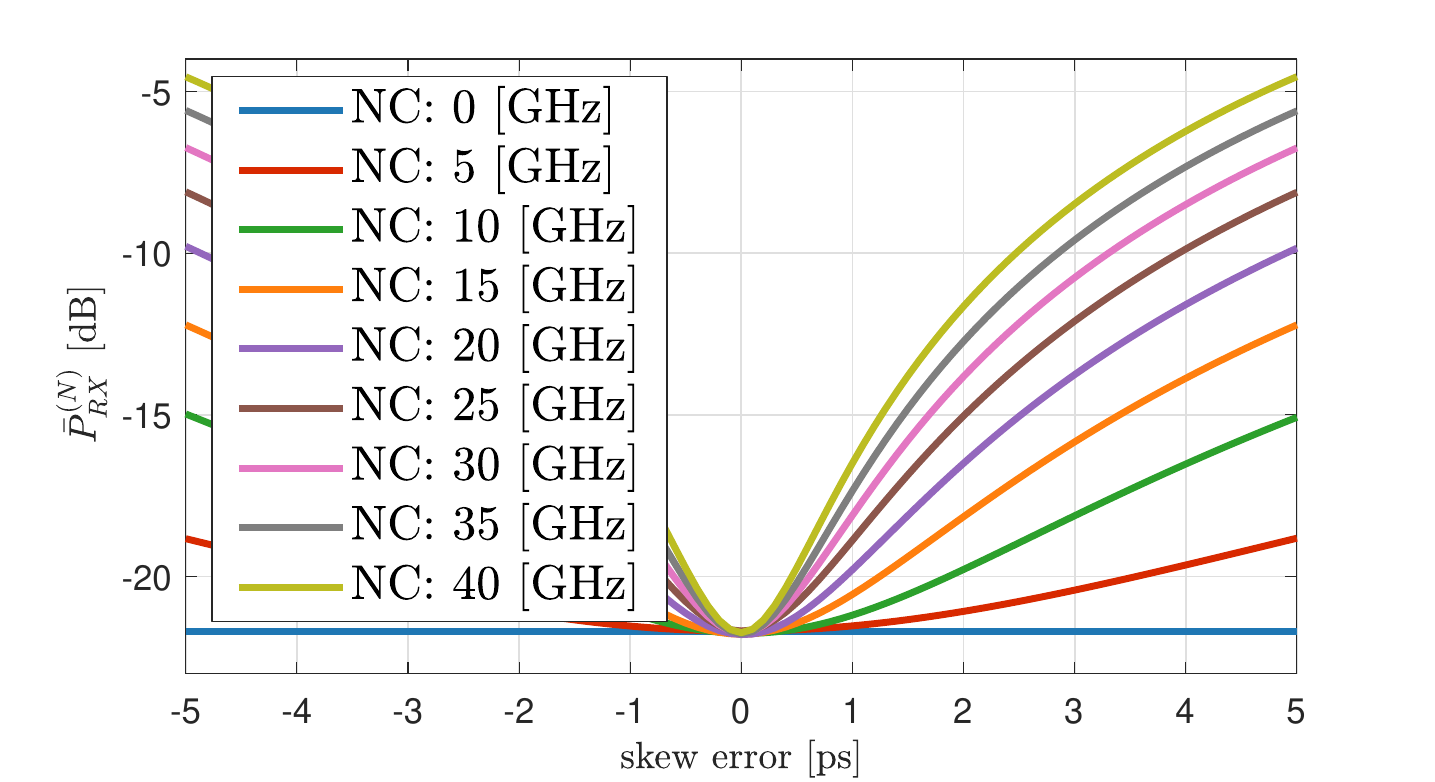}
\caption{Simulation of Evolution of the APSD of $F_{-N}$ as a function of skew and NC considering a -21 dB NFL.}
\label{fig:SkewErrorEvo}
\end{figure}

Given that the measured NFL can be affected by several impairments, the evolution of NFL as a function of the perturbation filters can be studied to determine uncompensated impairments. 

In this section, we demonstrate skew estimation in the x-polarization to illustrate a possible procedure in a Card2OSA configuration. This method can be generalized to a receiver accounting for both transmitted and receiver skew. The remaining of the impairments should be adequately compensated before the estimation of skew to avoid interactions between them.

Figure \ref{fig:PhaseEst} illustrates the principle of estimation, where the transmission of an engineered signal ($TX_2$) counteracts the NFL enhancement that takes place during the transmission through the linear system $H_{phase}(f)$. Hence, the minimization of the notch noise floor can be studied to estimate signal impairments and for diagnosis of calibration.

Two spectral regions are identified: a negative perturbation in the range $F_{N}$ whose gain is set to zero, and its symmetrical perturbed frequency range $F_{-N}$ whose amplitude is unperturbed. For the frequency range of $F_{-N}$, only the phase of the Q component is modified, given that:
\begin{equation}
\begin{split}
H_{phase}(f,\theta_k) = 
\begin{cases}
exp(j \theta_k(f)), & f \in (F_N  \cup  F_{-N}) \\
1, & \text{otherwise}\\
\end{cases}
\end{split}
\label{eq:PhaseEst}
\end{equation}

Hence, a compensation filter $H_{phase}(f,-\theta_k)$ can be applied to the transmitted notched signal to minimize the NFL. Given that $\theta_k$ is unknown, a sweep can be performed where the NFL is being monitored. 

Figure \ref{fig:SkewErrorEvo} illustrates the NFL evolution as a function of the skew and the $F_N$ center (NC) proving that the cost function to minimize for the case of skew is convex and it will converge to its minimum skew value. Given the evolution of the cost function in Figure \ref{fig:SkewErrorEvo}, the technique is more sensitive to skew when the notch is further away from the carrier frequency. The Average Power Spectral Density (APSD) $\overline{P}_{SIG}^{(rg)}$ nomenclature is used, defined as:
\begin{equation}
\overline{P}_{SIG}^{(rg)} 	=	\int_{F_{rg}} |SIG(f)|^2 \, df	\, / \int_{F_{rg}} \, df
\end{equation}

\begin{figure}
\centering
\includegraphics[width=9.5cm]{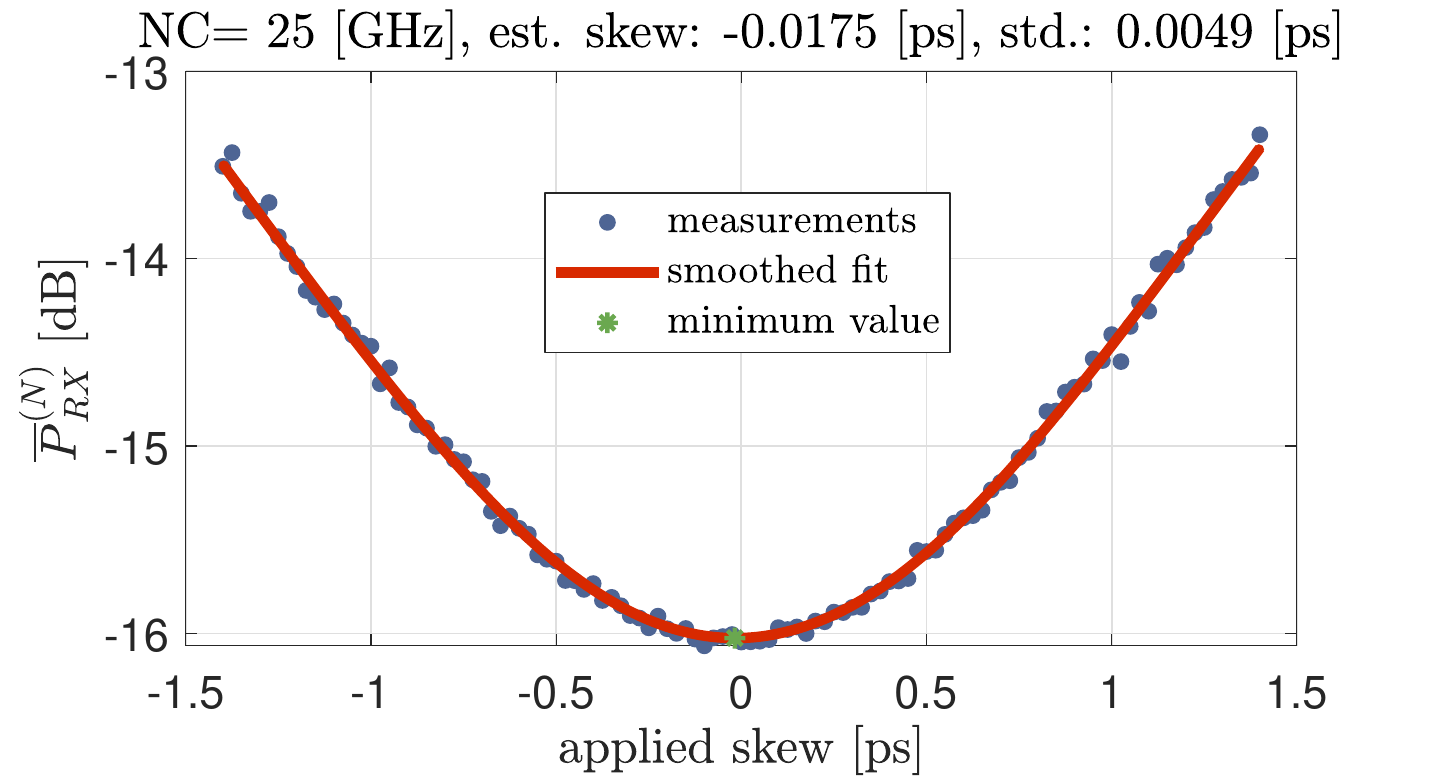}
\caption{Experimental Estimation of the skew for a Card2OSA configuration.}
\label{fig:SkewExpEst}
\end{figure}

Figure \ref{fig:SkewExpEst} experimentally estimates the skew for a Card2OSA scenario. For this experimental verification, the skew was varied between -1.4 to 1.4 ps in steps of 0.25 ps, where each measurement consisted of 40 Waveanalyzer traces of an x-polarization only transmission. 8 iterations were performed to estimate a standard deviation (std) of 0.049 ps showing consistent results and proving the repeatability of this approach.

\section{Conclusion}
In this paper, we investigate a frequency domain perturbation approach to measure frequency SNDR. The proposed technique is suitable for the estimation of noise contributions at different stages of optical transceivers and with different perturbation topologies.

Different noise contributions can be estimated depending on where the measurements are performed: transmitter electrical noise, transmitter noise, and transceiver noise until ADC.

Additionally, we compare single notch and dual notch topologies. The differences between single notches and dual notches detect uncompensated impairments such as frequency-dependent crosstalk, IQ imbalance, skew, timming misalignments, or uncompensated phase response, and ultimately it can be used to detect impairments in transceivers.

In this paper, we consider the estimation of skew based on perturbations. Our analysis shows that the skew's cost function is convex, and it is possible to minimize the skew by minimizing the APSD in the notch.
We also provide an example where the skew of a transmitter was estimated with a perturbative approach. The analysis of how the remaining impairments affect single notches, and the associated cost functions for their estimation will be the topic of future work.

\section*{Acknowledgments}
The authors would like to gratefully acknowledge discussions with D. Charlton, C. Laperle, M.E Mousa-Pasandi, M. Hubbard, M. Reimer, and M. O'Sullivan; and thank Ciena for the donation of equipment, funding, and support.

\ifCLASSOPTIONcaptionsoff
  \newpage
\fi

\bibliographystyle{ieeetr}
\bibliography{library}

\end{document}